\documentclass[%
 reprint,
 amsmath,amssymb,
aps
]{revtex4-1}
\usepackage{url}
\usepackage[colorlinks=true, linkcolor=blue,urlcolor=black, anchorcolor=blue, citecolor=blue, bookmarksnumbered]{hyperref}
\usepackage{graphicx}
\usepackage{dcolumn}
\usepackage{bm}
\usepackage[separate-uncertainty=true]{siunitx}
\usepackage{subfigure}
\usepackage{threeparttable}
\usepackage{booktabs}
\newcommand{\tabincell}[2]{\begin{tabular}{@{}#1@{}}#2\end{tabular}}

\begin{document}

\title{Prospective Study on Observations of $\gamma$-Ray Sources in the Galaxy Using the HADAR Experiment}

\author{
	Xiangli Qian$^{1,2}$,
	Huiying Sun$^1$,
	Tianlu Chen$^{2,\dag}$,
	Danzengluobu$^2$,
	Youliang Feng$^2$,
	Qi Gao$^2$,
	Quanbu Gou$^3$,
	Yiqing Guo$^{3,4,\ddag}$,
	Hongbo Hu$^{3,4}$,
	Mingming Kang$^5$,
	Haijin Li$^2$,
	Cheng Liu$^3$,
	Maoyuan Liu$^2$,
	Wei Liu$^3$,
	Bingqiang Qiao$^3$,
	Xu Wang$^1$,
	Zhen Wang$^6$,
	Guangguang Xin$^7$,
	Yuhua Yao$^5$,
	Qiang Yuan$^8$,
	Yi Zhang$^8$
}


\affiliation{
	$^{1}$ School of Intelligent Engineering, Shandong Management University, Jinan 250357, China\\
	$^{2}$ The Key Laboratory of Cosmic Rays (Tibet University), Ministry of Education, Lhasa 850000, China \\
	$^{3}$ Key Laboratory of Particle Astrophysics, Institute of High Energy Physics, Chinese Academy of Sciences, Beijing 100049, China \\
	$^{4}$ University of Chinese Academy of Sciences, 19 A Yuquan Road, Shijingshan District, Beijing 100049, China\\
	$^{5}$College of Physics, Sichuan University, Chengdu 610064, China\\
	$^{6}$Tsung-Dao Lee Institute, Shanghai Jiao Tong University, Shanghai 200240, China\\
	$^{7}$School of Physics and Technology, Wuhan University, Wuhan 430072, China\\
	$^{8}$Key Laboratory of Dark Matter and Space Astronomy, Purple Mountain Observatory, Chinese Academy of Sciences, Nanjing 210008, China
}
\affiliation{
	Corresponding author. E-mail:
	$^\dag$\href{mailto:chentl@ihep.ac.cn}{chentl@ihep.ac.cn}
	$^\ddag$\href{mailto:guoyq@ihep.ac.cn}{guoyq@ihep.ac.cn}	
}

\begin{abstract}
The High Altitude Detection of Astronomical Radiation (HADAR) experiment is a refracting terrestrial telescope array based on the atmospheric Cherenkov imaging technique. It focuses the Cherenkov light emitted by extensive air showers through a large aperture water-lens system for observing very-high-energy $\gamma$-rays and cosmic rays. With the advantages of a large field-of-view (FOV) and low energy threshold, the HADAR experiment operates in a large-scale sky scanning mode to observe galactic sources. This study presents the prospects of using the HADAR experiment for the sky survey of TeV $\gamma$-ray sources from TeVCat and provids a one-year survey of statistical significance. Results from the simulation show that a total of 23 galactic point sources, including five supernova remnant sources and superbubbles, four pulsar wind nebula sources, and 14 unidentified sources, were detected in the HADAR FOV with a significance greater than 5 standard deviations ($\sigma$). The statistical significance for the Crab Nebula during one year of operation reached \SI{346.0}{\sigma} and the one-year integral sensitivity of HADAR above \SI{1}{TeV} was $\sim$1.3\%--2.4\% of the flux from the Crab Nebula.
\end{abstract}
\maketitle

\renewcommand{\thefootnote}{$^{\arabic{footnote})}$}

\section{Introduction}
Very high energy (VHE; $\ge$ \SI{10}{\GeV}) $\gamma$-rays provide a fascinating insight into cosmic particle acceleration processes in the extreme environments of the local universe. Within our Galaxy, such high-energy emissions mainly originate from pulsars, pulsar wind nebulae (PWNe), supernova remnants (SNRs), binaries, and small quasars. Outside our galaxy, VHE $\gamma$-rays, mostly come from quasars and blazars, including BL Lac objects and flat-spectrum radio quasars. They are observed in extreme relativistic jets as particles escape from supermassive black holes. VHE gamma radiation is thought to primarily originate from the interactions of relativistic particles. Relativistic electrons produce $\gamma$-rays through non-thermal bremsstrahlung or inverse Compton scattering, whereas protons and atomic nuclei generate $\gamma$-rays through the decay of neutral $\pi^{0}$ mesons produced through hadronic interactions with ambient material. Therefore, VHE $\gamma$-ray astronomy provides the most direct means for exploring non-thermal astrophysical processes in the universe.

At TeV energies, supernovae and SNRs are generally considered the main sources of VHE $\gamma$-rays in the galaxy (E \textgreater \SI{e15}{\eV}). 
SNRs are assumed to be major cosmic-ray acceleration regions for the following two principal reasons: (1) According to the well-established theory of diffusive shock wave  acceleration~\cite{shcok acceleration 1983, paticle acceleration 1987}, projectiles resulting from supernova explosions may diffuse into the surrounding medium to form shock waves. The particles are accelerated at SNR shock wave fronts and the power index of the accelerated particle spectrum is estimated to be 2~\cite{TeV gamma astro 2012, Schure 2012}. This result is consistent with the $\gamma$-ray spectrum from supernova remnants, which is observed in the radio waveband. (2) About \SIrange[range-phrase = --,range-units = single]{e50}{e51}{ergs} of energy is released into cosmic space each time a supernovae explodes, which is the same as the rate of cosmic ray energy loss in the galaxy. Therefore, galactic supernovae and the resulting SNRs are the only potential sources that can provide the amount of energy required to produce the measured local cosmic-ray energy density~\cite{origin of CR 1964, yuan 2012}. Other potential sources of VHE $\gamma$-rays include pulsars and PWNs. Pulsars, which are produced by the explosion of Type Ib/Ic and II supernovae, are rapidly rotating magnetized neutron stars and act like unipolar inductors accelerating particles. They emit pulsar winds made up of electron-positron pairs, and the ultra-relativistic pulsar wind interacts with ambient supernova projectiles to create termination shocks. Here, electrons are accelerated, emitting TeV $\gamma$-rays via inverse Compton scattering. Thus, the radiative constituents of such celestial bodies tend to exhibit a pulsed radiation component from the immediate vicinity of the pulsar and an unpulsed component from the shock region and beyond. Therefore, a survey of the inner part of the galaxy provides an efficient approach to search for yet unknown types of galactic VHE $\gamma$-ray emitters and understand the origin and acceleration mechanism acting on them.

Over 200 known VHE $\gamma$-ray sources have been discovered in our galaxy~\cite{Hess survery 2018}. A large fraction of these sources were first discovered by the current generation of Imaging Atmospheric Cherenkov Telescopes (IACTs), namely H.E.S.S.~\cite{Hess project 2004}, MAGIC~\cite{magic partI 2016}, and VERITAS~\cite{veritas 2006}. IACTs are designed to detect the faint Cherenkov flashes originating from particle showers that form when energetic photons impacts the Earth's atmosphere. For all three of the telescope systems mentioned above, the Cherenkov light is focused into meter-size cameras through large reflective mirrors and then digitized through photomultiplier tubes (PMT). The typical signature of a $\gamma$-ray induced shower in the camera is an elliptic image over a duration of a few nanoseconds. In contrast, a proton-induced shower produces an irregular image, by which the $\gamma$-ray is distinguished from the dominating background. The next generation Cherenkov telescope array (CTA)~\cite{cta 2018} aims to observe $\gamma$-ray astronomy in the $\sim$\SI{20}{\GeV}--\SI{300}{\TeV} energy range. Possessing significantly superior sensitivity and angular resolution, it is the latest breakthrough in VHE $\gamma$-ray astronomy~\cite{hundreds tev 2015}. However, due to the influence of sky brightness and weather conditions, the duty cycle of current IACTs is below 10\% and the field-of-view (FOV) is small ($\sim$3.5--\ang{5}). Therefore, ICATs are more suitable for point source observation rather than all-sky scanning with a large FOV. On the contrary, for the extensive air shower (EAS) ground-based observatory array, the observation is not limited by the environment or weather conditions, although the sensitivity is relatively low. With a large FOV ($\sim$\SI{1}{sr}), EAS ground-based experiments observe the VHE $\gamma$-ray sources continuously, especially in the TeV energy range. So far, HAWC~\cite{HAWC 2014} has operated large-scale sky surveys of the Crab Nebula flux in the northern sky with a \SI{1}{yr} survey sensitivity of $\sim$5\%--10\%~\cite{HAWC 2017}. Milagro~\cite{Milagro survey 2004} and ARGO-YBJ~\cite{argo-ybj gamma survey 2013} have also performed similar surveys. The new complex EAS array LHAASO~\cite{LHAASO_sciencebook 2022} has detected photons with energies exceeding \SI{1}{PeV}, including one photon at \SI{1.4}{PeV}~\cite{1.4PeV 2021}. Besides, the Tibet AS$\gamma$~\cite{ASgamma 100TeV 2019} and HAWC~\cite{hawc 56TeV 2020} experiments have detected $\gamma$-rays beyond \SI{100}{\TeV} from several sources, which are very good candidates for PeVatrons. The new TAIGA-HiSCORE~\cite{taiga-hiscore 2017} non-imaging Cherenkov array experiment will have an exploration of $\gamma$-rays above \SI{30}{\TeV} and cosmic rays above \SI{100}{\TeV}. These observations enhance the knowledge of the origins and acceleration mechanism of VHE $\gamma$-ray emission. 

However, for violent explosions in the universe such as $\gamma$-ray bursts and supernova explosions, the duration of the transient sources is short. Although IACTs have adequate sensitivity and angular resolution, these transient sources cannot be captured immediately due to the limited FOV. As for EAS facilities, the energy threshold is relatively high (hundreds of GeV) and it has a large effective area and high duty cycle. However, it is limited to detecting only low-energy gamma emission. Based on this, the High Altitude Detection of Astronomical Radiation (HADAR), a ground-based experimental array with a low energy threshold ($\sim$\SI{10}{\GeV}) and a large FOV, was proposed and constructed. HADAR is a ground-based telescope array based on the atmospheric Cherenkov imaging technique, which detects the Cherenkov light through a large aperture wide-angle water-lens (lens and purified water) system for observing VHE cosmic rays and $\gamma$-rays. One of the scientific goals of HADAR is to accurately observe $\gamma$-ray sources (point sources and transient sources) from \SI{10}{\GeV} to tens of TeV. 

In 2016, a 0.9 meter diameter water-lens telescope prototype was successfully operated at the YangBaJing Cosmic Ray Observatory (\SI{4300}{\m} a.s.l., $90.522^{\circ}$\,E, $30.102^{\circ}$\,N, \SI{606}{g/cm^2}) in Tibet, China. It successfully detected a coincident cosmic ray event for the first time through the joint operation of adjacent scintillation detector array~\cite{caih 2017}. For a detailed description of the prototype performance, refer to~\cite{chentl 2019}. These results confirmed that this new detection technique was capable of imaging atmospheric Cherenkov light. Subsequently, the second step of the experimental design was implemented, where two or three \SI{1}{\m} diameter hemispheres were used as the lens body. To date, two lenses have been produced, the relevant upgraded electronics and data acquisition systems are being prepared, and installation and operation are scheduled for later this year. The plan for the third phase is to construct a HADAR telescope array with four \SI{5}{\m} diameter hemispherical lenses, to detect high-energy $\gamma$-ray emissions more effectively.

In this study, we present the prospects for a sky scanning survey of TeV $\gamma$-ray sources in the galaxy with HADAR. The paper is organized as follows: the HADAR experiment and its performance are introduced in Sect.~\ref{section2}, while Sect.~\ref{section3} describes the simulation and analysis method for the observation of sources. In Sect.~\ref{section4}, we report on the prospective observation results of different types of $\gamma$-ray sources, while Sect.~\ref{section5} presents a summary and discussion of this study.

\section{HADAR Experiment}\label{section2}
\begin{figure}[!htb]   
	\vspace*{1mm}   
	\centering    
	\subfigure{
		\label{fig1a}
		\includegraphics[width=0.7\hsize]{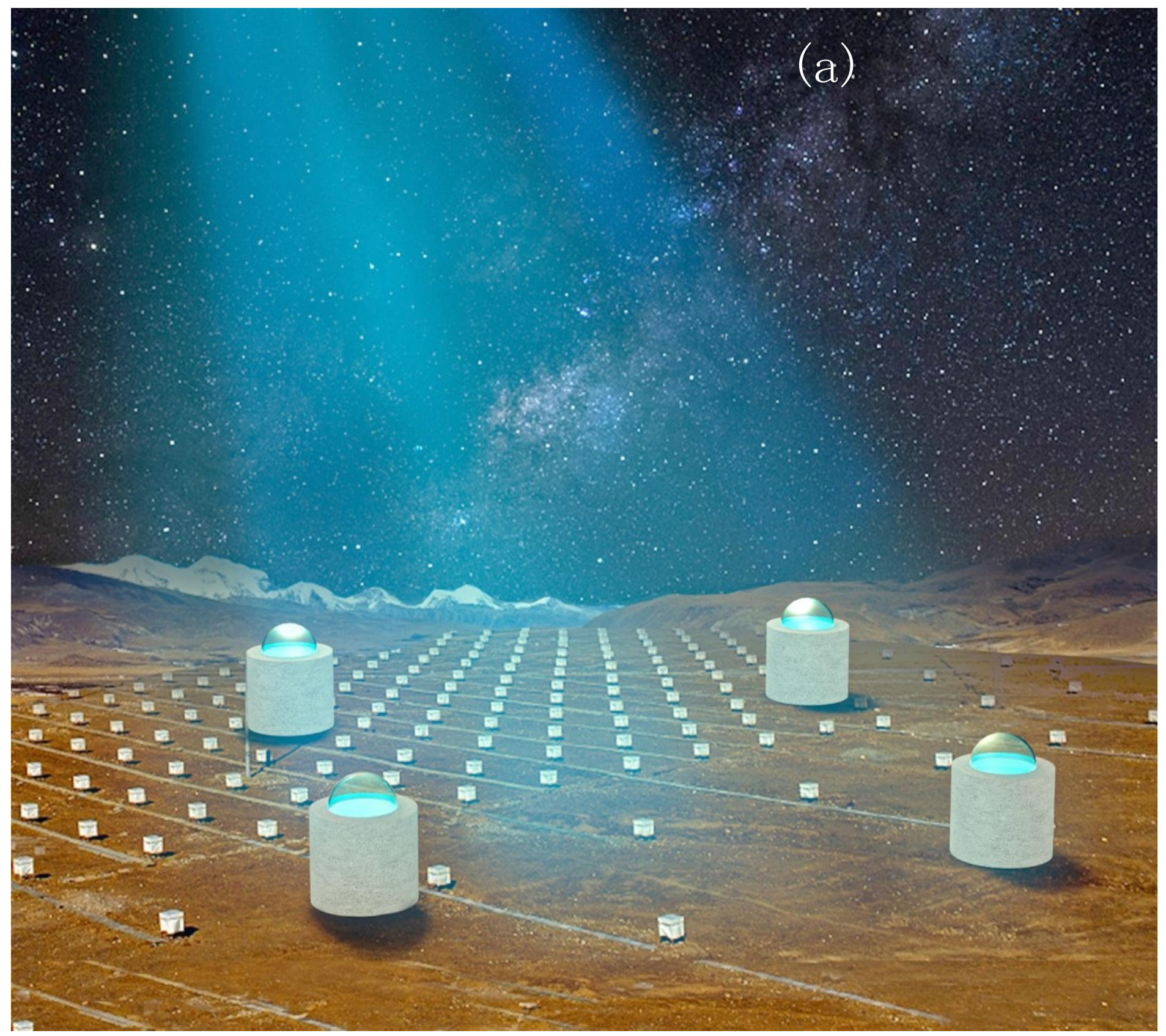}
	}
	\subfigure{
		\label{fig1b}
		\includegraphics[width=0.7\hsize]{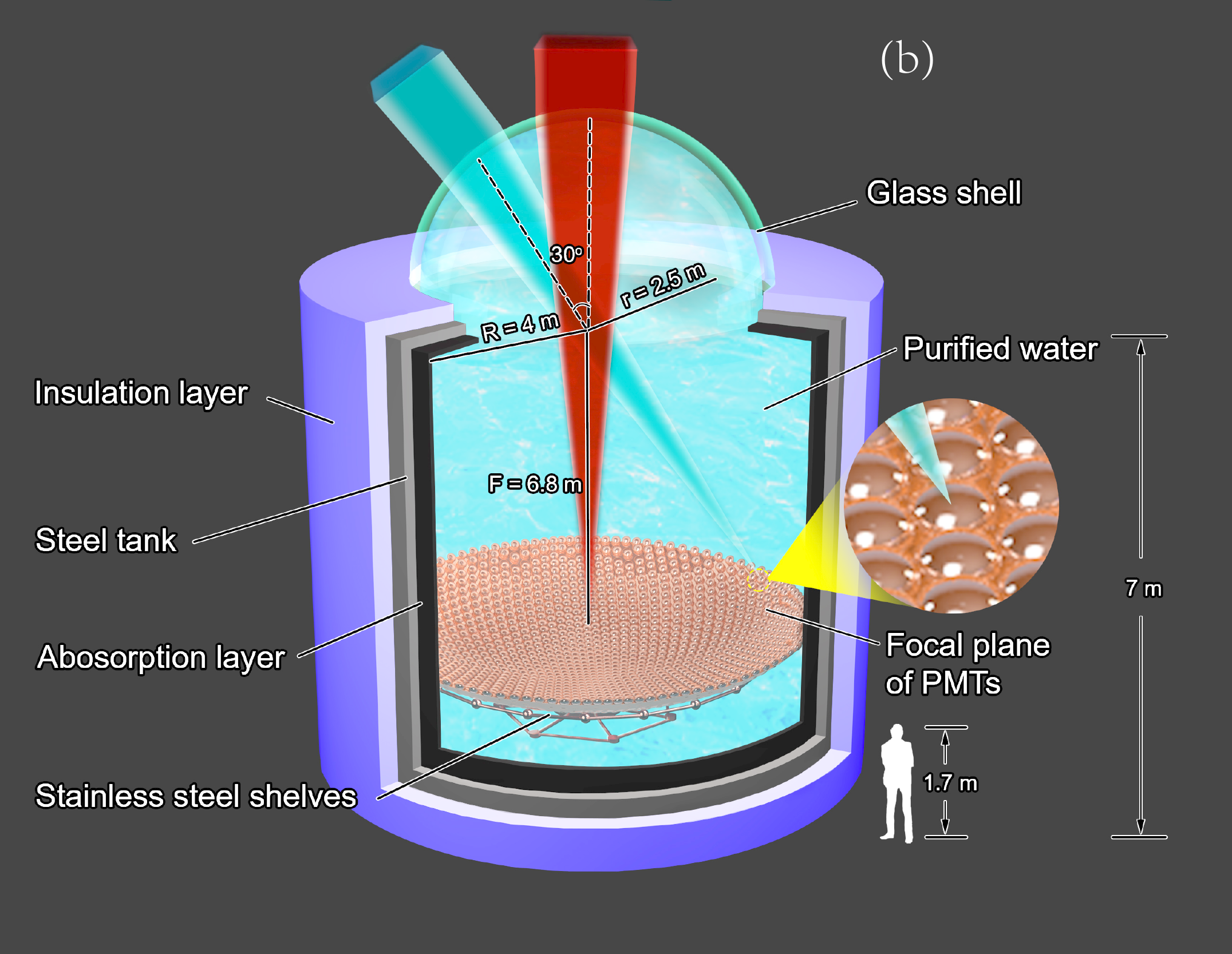}
	}
	\\[1mm] 
	\caption{Schematic of HADAR: \textbf{(a)} Layout of the HADAR experiment; \textbf{(b)} Detailed design of a water-lens telescope~\cite{xingg 2021}.}\label{fig1}
\end{figure}

\vspace*{1mm}
\noindent
The HADAR experiment is a hybrid array consisting of four water-lens telescopes and a surrounding scintillation detector array for observing Cherenkov light induced by \SI{10}{\GeV}--tens of TeV cosmic rays and $\gamma$-rays in the atmosphere. The schematic diagram of HADAR is shown in Fig.~\ref{fig1a}. The water-lens telescope array and the surrounding plastic scintillation array, i.e., the YangBaJing Hybrid Array~\cite{hybrid_ybj 2018}, were combined for observation. The four water-lenses were arranged in a square of \SI{100}{\m} spacing, considering that the radius of the Cherenkov light pool produced by a $\gamma$-ray shower with a primary energy of \SI{1}{\TeV} was about \SI{125}{\m}~\cite{gamma-ray telescope 2009}. Figure~\ref{fig1b} presents the detailed design of the water-lens telescope, which mainly consists of a hemispherical lens with a diameter of \SI{5}{\m} acting as a Cherenkov light collector, a cylindrical metal tank with a \SI{4}{\m} radius and \SI{7}{\m} height, and an imaging system at the bottom of the tank. The tank was filled with high-purity water to improve the transmissivity of ultraviolet photons. The inside of the tank was lined with an absorption layer, while an insulation layer was attached to the outside. The imaging system consisted of 18,961 2-inch PMTs arranged on the focal plane of the lens for image digitization. It was designed as a curved surface supported by a stainless-steel bracket for imaging the photons with large incident angles, and the focal length was \SI{6.8}{\m}. According to Fig.~\ref{fig1b}, green parallel light with an incident angle of \ang{30} was focused on the edge of the imaging system, allowing the FOV to reach \ang{60}. As a novel experiment, we adopted cheap acrylic and high purity water as the raw materials for the water-lens telescope, to keep the costs relatively low.

\begin{figure}[!htb]
	\includegraphics[width=0.9\hsize]{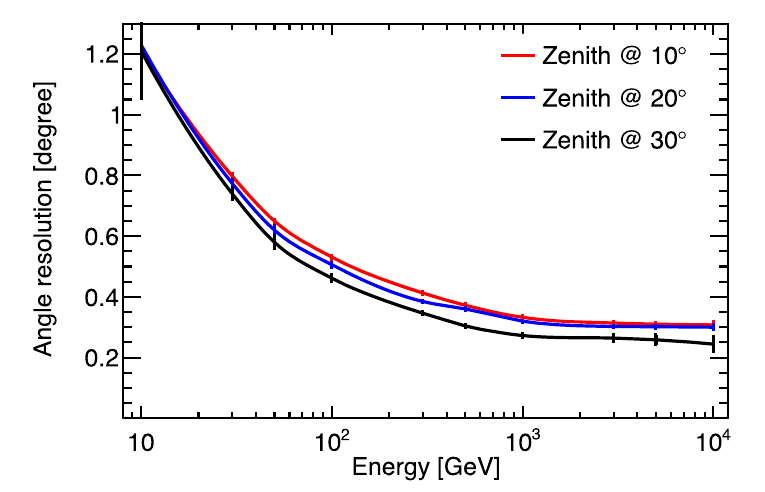}
	\caption{Performance of HADAR angular resolution. The incident zenith angles were \SIlist{10;20;30}{\degree}~\cite{xingg 2021}.}
	\label{fig2}
\end{figure}

\begin{figure}[!htb]  
	\subfigure{
		\label{fig3a}
		\includegraphics[width=0.95\hsize]{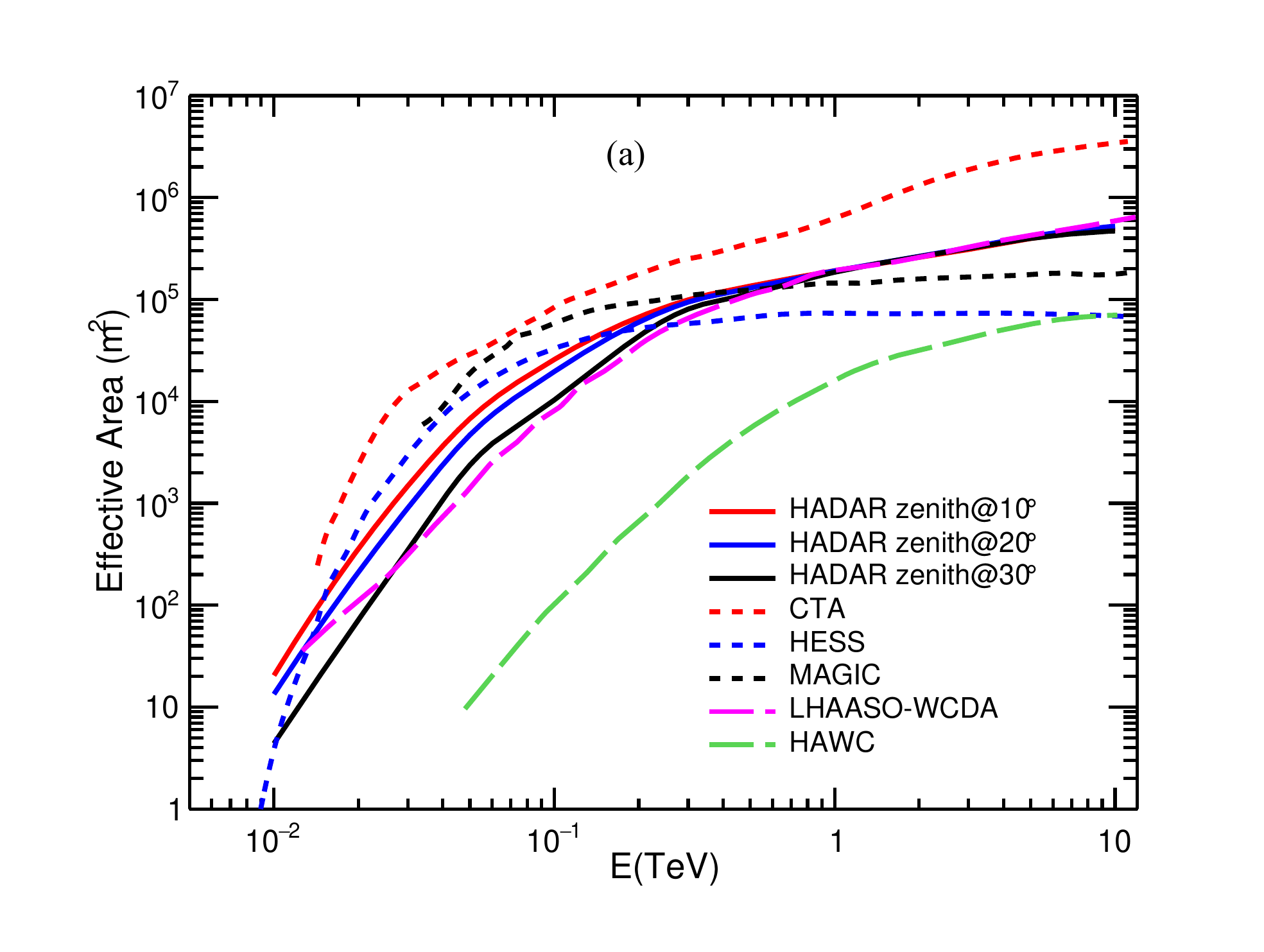}
	}
	\subfigure{
		\label{fig3b}
		\includegraphics[width=0.95\hsize]{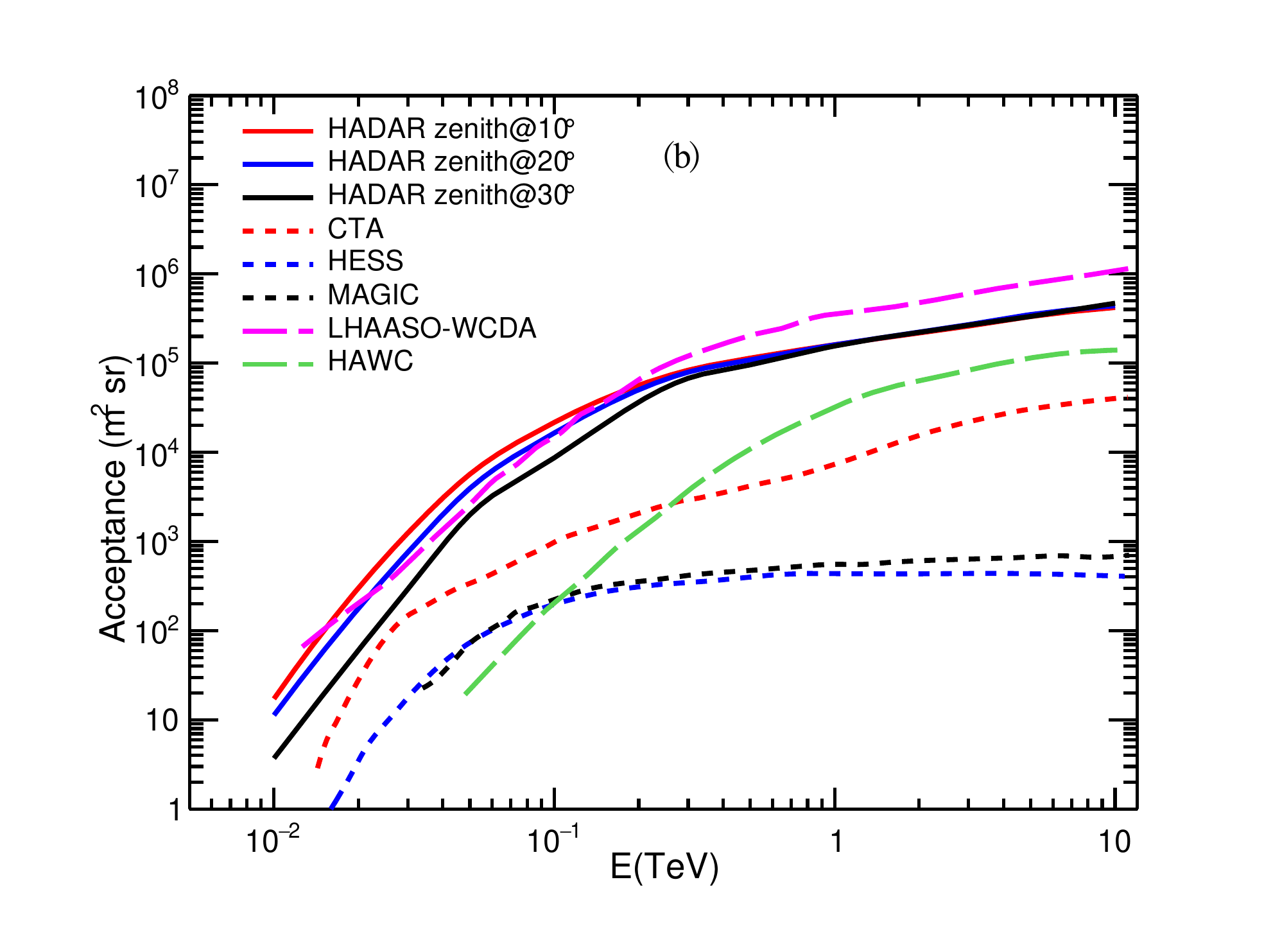}
	}
	\caption{Performance of HADAR effective area: \textbf{(a)} Effective areas for incident $\gamma$-ray events at different zenith angles~\cite{xingg 2021} and comparison with other IACT and EAS experiments; \textbf{(b)} Acceptance for HADAR and other experiments.}
	\label{fig3}
\end{figure}

Figure~\ref{fig2} displays the angular resolution (point-spread function, PSF; 68\% containment radius) of HADAR for incident $\gamma$-ray events at different zenith angles. The incident zenith angles were \SIlist{10;20;30}{\degree}, respectively, and the energy ranged from \SI{10}{\GeV} to \SI{10}{\TeV}. The angular resolution was about a subdegree and was approximately \SI{0.4}{\degree} at $\sim$\SI{300}{\GeV}. Figure~\ref{fig3} illustrates the effective areas of HADAR as well as a comparison with other IACT experiments. The effective area increased from about \SI{10}{\m^2} to \SI{4e5}{\m^2} as the energy increased, as indicated by Fig.~\ref{fig3a}. Also, the effective area was approximately \SI{1e5}{\m^2} at $\sim$\SI{300}{\GeV} for HADAR, which was comparative to the values generated by H.E.S.S.~\cite{hess photon 2015}, MAGIC~\cite{magic partII 2016} and LHAASO~\cite{LHAASO_sciencebook 2022}, and much larger than HAWC~\cite{HAWC effective area}. The effective area of HADAR was inferior to that of IACTs in the low energy range ($\textless$ \SI{300}{\GeV}) because the diameter of the HADAR telescope was only \SI{5}{\m} and the number of Cherenkov lights collected was relatively small. Regarding the incident events of different zenith angles, there was little difference in effective area. Figure~\ref{fig3b} provides a comparison of the acceptance (the acceptance is defined here as the integral of the effective area over the FOV) in different experiments. The region covered by HADAR was much larger than other IACTs (even the CTA), and comparable to LHAASO. The key performance parameters of HADAR are shown in Table~\ref{tab1}. Owing to its merits, HADAR is suitable for searching for $\gamma$-ray point and extended sources in the northern sky. Additionally, HADAR also has the advantage of observing high-energy $\gamma$-ray bursts, and the reference~\cite{xingg 2021} shows expectations for the observations.

\begin{table}[!htb]
	\caption{Performance of the HADAR experiment. The spatial coverage, energy range, effective area, angular resolution, energy resolution, and point-source sensitivity are given. For a detailed description of the flux sensitivity, see Sect.~\ref{section4}.}
	\label{tab1}
	\tabcolsep 6.0pt
	\renewcommand{\arraystretch}{1.2} 
	\begin{tabular*}{8.5cm}{cc} 
		\toprule
		\hline
		Performance & Values \\ \hline		
		Field of view (sr) & 0.84 \\ 	 
		Sky Coverage (dec.) &  0.102--\ang{60.102}\\
		Energy & \SI{30}{\GeV}--\SI{10}{\TeV} \\
		Effective Area (\SI{100}{\GeV}) & \SI{2.4e5}{\m^2}\\
		Angular Resolution (\SI{100}{\GeV}) & \SI{0.44}{\degree}\\
		Energy Resolution (\SI{100}{\GeV}) & 30\%\\ 
		Sensitivity (Crab) & $\sim$1.3--2.4\%\\
		\bottomrule
		\hline
	\end{tabular*}
\end{table}

\section{Expectations on Observations of $\gamma$-Ray Sources Using HADAR}\label{section3}
Cosmic ray events account for the vast majority of detected events when tracing and counting the number of events collected from a point source direction. Therefore, the significance of observing galactic TeVCat sources is to accurately estimate the number of cosmic ray background events and the $\gamma$-ray events from each source. For this, we performed a fast simulation to generate $\gamma$-ray and cosmic ray events and adopted a sky scanning analysis method based on the equi-zenith angle~\cite{asgamma 2005} to calculate the significance. Based on the measured energy spectrum data of TeVCat sources provided by previous experiments, we estimated the detection capability of HADAR for these sources and calculated the expected statistical significance of a \SI{1}{yr} sky survey. The energy for the simulation ranged from \SI{30}{GeV} to \SI{10}{TeV}, with a zenith angle of less than \ang{30}. 

\subsection{Simulation Method for Observing $\gamma$-Ray sources}\label{sec3.1}
Considering the zenith angle range and the location of the HADAR experiment in the horizontal coordinate system (observer's coordinates), the sky map in the equatorial coordinate was divided into \ang{0.1}$\times$\ang{0.1} bin pixels, from \ang{0} to \ang{360} in right ascension and from \ang{0.102} to \ang{60.102} in declination. Each pixel was denoted as $(i, j)$ and the center of a pixel represented the location of the point source. Assuming a local sidereal time bin (t), the FOV of HADAR in the horizontal coordinates was divided into windows according to the following criteria: the zenith angle $\theta$ was divided into \ang{0.08} bins from \ang{0} to \ang{30}, and the azimuth $\phi$ was divided into $\frac{0.08^{\circ}}{\sin\theta}$ bins from \ang{0} to \ang{360}, for which the width depended on the zenith angle. Thus, the solid angle for each azimuth window was approximately equal, $\Omega=\num{1.95e-6}$, and each window was numbered $(t, \theta, \phi)$. 

In the horizontal coordinate system, the depths for cosmic rays with different zenith angles traveling through the atmosphere vary. As a result, the detection efficiency decreases with an increase in zenith angle, while the atmospheric depths at different azimuths under the same zenith angle remain the same. Therefore, it is more reasonable to estimate the cosmic ray background for a point source using collected shower events in the same zenith angle belt and different azimuth angles from the signal window. For a zenith angle belt, the window pointing to the source is known as the "on-source window", while the sideband windows on the same belt are called "off-source windows". Thus, the background events for the "on-source window" can be estimated as the average value of all events in the "off-source windows".  Due to the rotation of the Earth, the $\gamma$-ray source signals detected by HADAR form a trajectory within windows $(\theta, \phi)$ in the horizontal coordinates. Each window in one time bin $(t, \theta, \phi)$ is always mapped to a unique pixel $(i, j)$ in the equatorial coordinates, that is, $(t, \theta, \phi) \to (i, j)$. The event numbers collected in each window are denoted as $N(t, \theta, \phi)$, thus the event number in pixel $(i, j)$ in the equatorial coordinates is a summation of the number of events counted in the horizontal coordinate, i.e., $N(i, j)= \sum N_{\rm obs}(t, \theta, \phi)$. The relative intensity of the signals in the "on-source window" is defined as $I(i, j)$, which is the ratio of the number of events observed in the "on-source window" to the number of background events, i.e., $N_{\rm on}(t, \theta, \phi)/N_{\rm off}(t, \theta, \phi)$. Therefore, the number of normalized events relative to $I(i, j)$ in the "on-source window" is $N_{\rm obs}(t, \theta, \phi)/I(i, j)$. Similarly, the "off-source window" with the same zenith angle ($\theta$) is expressed as $(t, \theta, \phi^{'})$, with corresponding coordinates $(i^{'}, j^{'})$ in the equatorial coordinates. The number of events observed in the "off-source window" is $N_{\rm obs}(t, \theta, \phi^{'})$, the relative intensity is $I(i^{'}, j^{'})$, the number of normalized events is $N_{\rm obs}(t, \theta, \phi^{'})/I(i^{'}, j^{'})$, and the average number of normalized events in all "off-source windows" is $\frac{1}{n_{\theta}-1}\sum_{\phi^{'}}\frac{N_{\rm obs}(t, \theta, \phi^{'})}{I(i^{'}, j^{'})}$, where $n_{\theta}$ is the number of azimuth windows in one zenith angle ($\theta$) belt.

For the time bin, a sidereal day is divided into 3,600 time bins, i.e., the time interval is 24 seconds. For HADAR, Cherenkov light can only be detected on moonless nights. Therefore, the effects of sunlight and moonlight are excluded when calculating the effective observation time. When the zenith angle of the sun and the moon in the horizontal coordinates is greater than \ang{90}, it is considered a moonless night for simulations. Besides, other environmental factors like thunderstorms affect observations, but we ignored these variables in the simulation. Due to the movement of sun, the moon, and the Earth, the value of $(\theta, \phi)$ in the horizontal coordinate varies every day for each $\gamma$-ray source. Therefore, the effective observation time over \SI{1}{yr} of operation is the sum of the observation times for each day.

\subsection{Significance Analysis Method}\label{sec3.2}
According to the location, direction, and energy spectrum information of the source, we simulated the observed events generated from each source. The $\gamma$-ray excess events at each pixel $(i, j)$ were calculated in combination with the estimated cosmic ray background events for the source region. Thus, we obtained the statistical significance of the source. The predicted number of cosmic rays in a window $(t, \theta, \phi)$ is determined as follows:
\begin{equation}\label{eq1}
N_{CR}(t, \theta, \phi) = \eta_{CR}\int_{E}J_{CR}(E)A_{CR}(\theta, E){\Omega}dE{\delta}t
\end{equation}
Besides, the predicted number of $\gamma$-ray events from the point source in a window $(t, \theta, \phi)$ is calculated as follows:
\begin{equation}\label{eq2}
N_{\gamma}(t, \theta, \phi) = \eta_{\gamma}\int_{E}J_{\gamma}(E)A_{\gamma}(\theta, E)dE{\delta}t{\varepsilon_{\gamma}}(\Omega)
\end{equation}
where $A_{CR}(\theta, E)$ and $A_{\gamma}(\theta, E)$ represent the effective areas of HADAR for the cosmic rays and $\gamma$-rays, respectively. Also, $J_{CR}(E)$ signifies the cosmic ray energy spectrum~\cite{cosmic ray spectrum 2013}, $J_{\gamma}(E)$ is the $\gamma$-ray energy spectrum for each source, $\Omega$ denotes the solid angle corresponding to the HADAR angular resolution, $\varepsilon_{\gamma}(\Omega)$ is 68\% and represents the containment percentage of $\gamma$-ray events in the solid angle $\Omega$, $\eta_{CR}$ and $\eta_{\gamma}$ signify the efficiencies of simulated cosmic rays and $\gamma$-rays according to the $\rm \gamma$/Proton separation, and ${\delta}t$ is the time bin, which is 24 seconds. The differential flux energy spectrum of the $\gamma$-ray source is described by a power law with an exponential cut-off, i.e.:
\begin{equation}\label{eq3}
\frac{dN}{dE}=J_0(\frac{E}{E_0})^{-\Gamma}e^{-\frac{E}{E_{cut}}}
\end{equation}
where $J_0$ represents the normalizaiton flux in units of \si{cm^{-2}.s^{-1}.TeV^{-1}} at energy $E_0$, $\Gamma$ is the energy spectrum index, and $E_{cut}$ denotes the cut-off energy for the source. For the specific information source concerning the energy spectrum, see Tables~\ref{tab2},~\ref{tab3} and~\ref{tab4}.

Statistically, the number of normalized events in the "on-source window" is equal to the average number of normalized events in the "off-source windows", i.e., $\frac{N_{obs}(t, \theta, \phi)}{I(i, j)} = <\frac{N_{obs}(t, \theta, \phi^{'})}{I(i^{'}, j^{'})}> $, while $\chi^2$ is determined according to the least square principle:
\begin{equation}\label{eq4}
\begin{aligned}
&\rm \chi^2=\sum_{t, \theta, \phi}\left\{{\left[\frac{N_{obs}(t, \theta, \phi)}{I(i, j)}-\frac{1}{n_{\theta}-1}\sum_{\phi^{'}}\frac{N_{obs}(t, \theta, \phi^{'})}{I(i^{'}, j^{'})}\right]}\right.\\
&\left.{\times\sigma_{t, \theta, \phi}^{-1}}\right\}^2
\end{aligned}
\end{equation}
By minimizing $\chi^2$, the relative intensity $I(i, j)$ and the corresponding statistical error $\Delta I(i, j)$ are obtained. Thus, the excess events $N_{\rm s}(i, j)$ and the uncertainty $\Delta N_s(i, j)$ in pixel $(i, j)$ in equatorial coordinates can be expressed as:
\begin{equation}\label{eq5}
N_s(i, j)=\frac{[I(i, j)-1]N(i, j)}{I(i, j)}
\end{equation}
\begin{equation}\label{eq6}
\Delta N_s(i, j)=\frac{\Delta I(i, j)N(i, j)}{I(i, j)}
\end{equation}

Considering the angular resolution of HADAR, signals from the point source may be reconstructed and subsequently deviate from the source direction in the reconstruction. To maximize the significance of the calculation, the smoothed $N_{\rm s}$ and $\Delta N_s$ values are used to approximately estimate the significance. The smoothed significance is calculated as the average significance of the region with HADAR angular resolution as the angular radius around the source center, thus $s = \frac{N_s}{\Delta N_s}$. 

Except for point sources, some sources have a certain extended morphology. Usually, a two-dimensional Gaussian model is convolved with the experimental point-spread function (PSF) to fit the excess map of the source~\cite{Hess source 2008}. Therefore, we use a two-dimensional Gaussian function:
\begin{equation}\label{eq7} 
f(x, y) = \frac{1}{2\pi{\sigma}^2}exp\left\{-\frac{1}{2}\left[\frac{(x-x_0)^2}{{\sigma_x}^2}+\frac{(y-y_0)^2}{{\sigma_y}^2}\right]\right\}
\end{equation}
This function simulates the $\gamma$-ray source extended emission positions in the sky map. The distribution of the extended emission is assumed to be rotationally symmetrical, i.e., $\sigma_x = \sigma_y = \sigma$, where $x_0$ and $y_0$ represent the $\gamma$-ray source positions in the equatorial coordinates. Details of the extended emissions for each source are listed in Tables~\ref{tab2},~\ref{tab3} and~\ref{tab4}.

\begin{table*}[!htb]
	\footnotesize
	\caption{Expected significance of SNRs, superbubbles, and binaries with HADAR between 30 GeV and 10 TeV using a \SI{1}{yr} observation time. Columns from left to right are as follows: source name, R.A., Dec., normalization flux, $\rm E_0$, spectral index, extended degree ("--" represents an extension of 0), effective livetime, expected significance by HADAR, and the references of the measurements.}
	\label{tab2}
	\tabcolsep 6.0pt
	\renewcommand{\arraystretch}{1.3} 
	\begin{tabular*}{\textwidth}{ccccccccccc}
		\toprule
		\hline
		Source & R.A.($^{\circ}$) & Dec.($^{\circ}$) & $J_0$ ($\rm TeV^{-1}cm^{-2}s^{-1}$)& $E_0$ (TeV) & $\Gamma$ & Extension ($^{\circ}$) & Livetime (hrs) & S($\sigma$) & Ref.\\ 
		\hline
		W 49B      		& 287.78 & 9.16	 & $\rm 3.15 \times 10^{-13}$ & 1 & 3.14 & --	& 195.0 & 8.6 & \cite{Hess survery 2018}\\
		HESS J1912+101	& 288.20 & 10.15 & $\rm 3.66 \times 10^{-14}$ & 7 & 2.64 & 0.7 & 203.3 & 20.6 &  \cite{HAWC 2017}\\
		W 51 			& 290.73 & 14.19 & $\rm 2.61 \times 10^{-14}$ & 7 & 2.51 & 0.9 & 230.4 & 8.2 & \cite{HAWC 2017}\\
		ARGO J2031+4157 & 307.80 & 42.50 & $\rm 3.50 \times 10^{-9}$  & 0.1 & 2.16 & 2.0 & 277.7 & 23.6 & \cite{J2031+4157}\\
		Cassiopeia A 	& 350.81 & 58.81 & $\rm 1.10 \times 10^{-11}$ & 0.433 & 2.40 & -- & 98.7 & 5.7 & \cite{cassiopeia cut-off 2017}\\
		\bottomrule
		\hline
	\end{tabular*}
\end{table*}

\begin{table*}[!htb]
	\footnotesize
	\caption{Expected significance of PWNs with HADAR between 30 GeV and 10 TeV using a \SI{1}{yr} observation time. Columns from left to right are as follows: source name, R.A., Dec., normalization flux, $\rm E_0$, spectral index, extended degree ("--" represents an extension of 0), effective livetime, expected significance by HADAR, and the references of the measurements.}
	\label{tab3}
	\tabcolsep 6.0pt
	\renewcommand{\arraystretch}{1.3} 
	\begin{tabular*}{\textwidth}{ccccccccccc}
		\toprule
		\hline
		Source & R.A.($^{\circ}$) & Dec.($^{\circ}$) & $J_0$ ($\rm TeV^{-1}cm^{-2}s^{-1}$)& $E_0$ (TeV) & $\Gamma$ & Extension ($^{\circ}$) & Livetime (hrs) & S($\sigma$) & Ref.\\ \hline
		Crab      		& 83.63  & 22.01 & $\rm 1.85 \times 10^{-13}$ & 7 & 2.58 & --  & 241.8 & 346.0 & \cite{HAWC 2017} \\
		Geminga	   		& 98.12  & 17.37 & $\rm 4.87 \times 10^{-14}$ & 7 & 2.23 & 2.0 & 230.2 & 13.7  & \cite{HAWC 2017} \\
		TeV J1930+188 	& 292.63 & 18.87 & $\rm 1.96 \times 10^{-14}$  & 7 & 2.18 & -- & 255.0 & 10.5  & \cite{VERITAS Fermi HAWC 2018} \\
		2HWC J1953+294  & 298.26 & 29.48 & $\rm 8.30 \times 10^{-15}$  & 7 & 2.78 & -- & 288.6 & 31.4  & \cite{HAWC 2017} \\
		\bottomrule
		\hline
	\end{tabular*}
\end{table*}

\begin{table*}[!htb]
	\footnotesize
	\caption{Expected significance of unidentified sources with HADAR between 30 GeV and 10 TeV using  a \SI{1}{yr} observation time. Columns from left to right are as follows: source name, R.A., Dec., normalization flux, $\rm E_0$, spectral index, extended degree ("--" represents an extension of 0), effective livetime, expected significance by HADAR, and the references of the measurements.}
	\label{tab4}
	\tabcolsep 5.3pt
	\renewcommand{\arraystretch}{1.3} 
	\begin{tabular*}{\textwidth}{ccccccccccc}
		\toprule
		\hline
		Source & R.A.($^{\circ}$) & Dec.($^{\circ}$) & $J_0$ ($\rm TeV^{-1}cm^{-2}s^{-1}$)& $E_0$ (TeV) & $\Gamma$ & Extension ($^{\circ}$) & Livetime (hrs) & S($\sigma$) & Ref.\\ \hline
		MAGIC J0223+403 & 35.80  & 43.01 & $\rm 1.70 \times 10^{-11}$ & 0.3 & 3.10 & --  & 252.5 & 13.5 & \cite{3C 66A/B MAGIC} \\
		2HWC J1829+070	& 277.34  & 7.03 & $\rm 8.10 \times 10^{-15}$ & 7 & 2.69 & -- & 173.8 & 14.4  & \cite{HAWC 2017} \\
		2HWC J1852+$\rm 013^*$ 	& 283.01 & 1.38 & $\rm 1.82 \times 10^{-14}$  & 7 & 2.90 & -- & 73.6 & 36.9  & \cite{HAWC 2017} \\
		MAGIC J1857.6+0297  & 284.40 & 2.97 & $\rm 6.10 \times 10^{-12}$  & 1 & 2.39 & 0.1 & 112.2 & 25.5  & \cite{Hess source 2008} \\
		2HWC J1902+$\rm 048^*$	& 285.51  & 4.86 & $\rm 8.30 \times 10^{-15}$ & 7 & 3.22 & -- & 145.7 & 82.1  & \cite{HAWC 2017} \\
		2HWC J1907+$\rm 084^*$	& 286.79  & 8.50 & $\rm 7.30 \times 10^{-15}$ & 7 & 3.25 & -- & 189.4 & 102.5  & \cite{HAWC 2017} \\
		MGRO J1908+06	& 286.98  & 6.27 & $\rm 8.51 \times 10^{-14}$ & 7 & 2.33 & 0.8 & 164.6 & 12.4  & \cite{HAWC 2017} \\
		2HWC J1914+$\rm 117^*$ & 288.68  & 11.72 & $\rm 8.50 \times 10^{-15}$ & 7 & 2.83 & -- & 215.3 & 29.1  & \cite{HAWC 2017} \\
		2HWC J1921+131	& 290.30  & 13.13 & $\rm 7.90 \times 10^{-15}$ & 7 & 2.75 & -- & 223.8 & 21.5  & \cite{HAWC 2017} \\
		2HWC J1928+177	& 292.15  & 17.78 & $\rm 1.07 \times 10^{-14}$ & 7 & 2.60 & -- & 249.9 & 19.8  & \cite{J1928+177 HAWC HESS} \\
		2HWC J1938+238	& 294.74  & 23.81 & $\rm 7.40 \times 10^{-15}$ & 7 & 2.96 & -- & 274.7 & 51.9  & \cite{HAWC 2017} \\
		2HWC J1955+285	& 298.83  & 28.59 & $\rm 5.70 \times 10^{-15}$ & 7 & 2.40 & -- & 286.7 & 7.3  & \cite{HAWC 2017} \\
		2HWC J2006+341	& 301.55  & 34.18 & $\rm 9.60 \times 10^{-15}$ & 7 & 2.64 & -- & 292.0 & 42.7  & \cite{HAWC 2017} \\
		VER J2019+407	& 305.02  & 40.76 & $\rm 1.50 \times 10^{-12}$ & 1 & 2.37 & 0.23 & 284.3 & 14.9  & \cite{VER J2019+407} \\	
		\bottomrule
		\hline
	\end{tabular*}
\end{table*}

\section{Expected Significance Results of $\gamma$-Ray Sources}\label{section4}
\begin{figure}[!htb]
	\includegraphics[width=1.0\hsize]{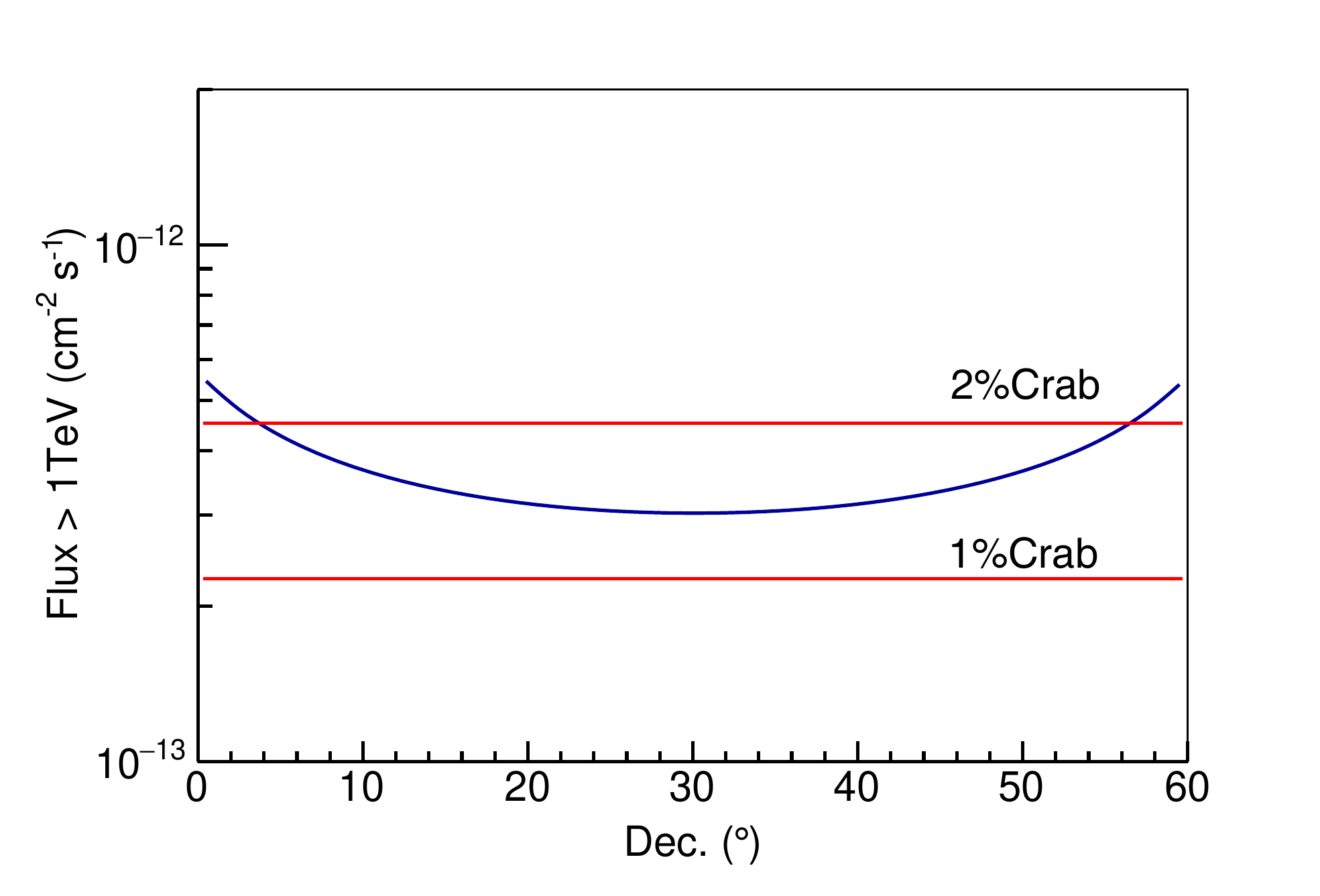}
	\caption{Sensitivity in equatorial coordinates for a one-year sky survey using HADAR in units of the flux from the Crab Nebula ($\sim$\SI{2.26e-11}{cm^{-2}.s^{-1}}, E $\textgreater$ \SI{1}{TeV})~\cite{crab hess 2006} and assuming a power-law spectrum with photon index $\Gamma = 2.51$. The sensitivity used here is defined as the minimum flux of a point source that would be detectable at the \SI{5}{\sigma} significance level in the HADAR FOV.}
	\label{fig4}
\end{figure}

Figure~\ref{fig4} displays the simulated observation sensitivity in equatorial coordinates for a one-year HADAR sky survey. It shows that the effective observation time of each point source varied with the declination. In the simulation, the flux from the Crab Nebula was used as the standard and the spectral index was 2.51. In the HADAR FOV, the flux integral sensitivity above \SI{1}{TeV} was between 1.3\% and 2.4\% of the flux from the Crab Nebula, and the most sensitive observation position was at \ang{30.1} declination, which corresponded to the latitude of HADAR.

\begin{figure}[!htb]
	\includegraphics[width=1.0\hsize]{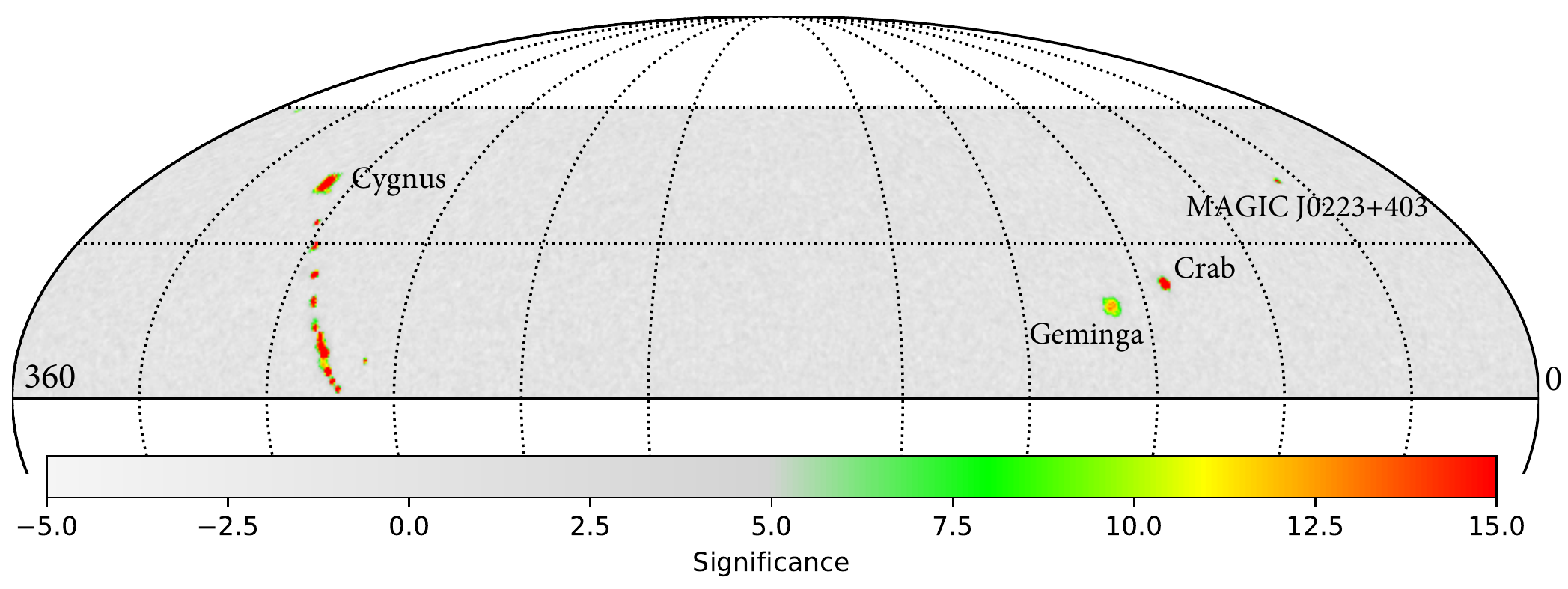}
	\caption{Expected significance of all TeV sources in the equatorial coordinates (J2000 epoch) in the HADAR FOV. The visualization range is limited between \num{-5} and 15.}
	\label{fig5}
\end{figure}

The expected results of sky survey observations from TeVCat sources in the galaxy by HADAR are shown in Tables~\ref{tab2},~\ref{tab3} and~\ref{tab4}, which display the details of each source (source name, location, normalization flux at energy $E_0$, spectral index, extended degree, effective livetime, and expected significance). A total of 23 galactic sources with significance greater than 5 standard deviations were observed in the HADAR FOV, including five SNRs and superbubbles, four PWNs, and 14 unidentified sources. Also, a two-dimensional skymap is presented in Fig~\ref{fig5}. Although the significance of many sources is greater than 15, the visualization range is confined between \num{-5} to 15. We list and briefly discuss the observations and studies of some sources by previous experiments in Sect.~\ref{4.1}--Sect.~\ref{4.4}, including their associated counterparts, flux, spectral information and so on. We also analyzed the potential of HADAR for sky surveys in the GeV energy band (Sect.~\ref{4.5}) and discussed the advantage of observing the extended sources using HADAR (Sect.~\ref{4.6}).\\

\subsection{Expected Observations of SNRs and Superbubbles}\label{4.1}

1. The expected significance of W49B (HESS J1911+090) observed by HADAR was \SI{8.6}{\sigma}. The SNR W49B is a mixed-morphology remnant originating from a core-collapse supernova that occurred between one and four thousand years ago. It is associated with SNR G43.3-0.2 and is believed to result from interaction with molecular clouds, exhibiting a soft energy spectrum. The H.E.S.S. collaboration reported that the integral flux above \SI{1}{\TeV} is  ($\rm 1.47\,\pm\,0.38_{stat}\,\pm\,0.29_{sys}) \times 10^{-13}$\,\si{cm^{-2}.s^{-1}} with a spectral index of \num{3.14}~\cite{Hess survery 2018}. H.E.S.S. alos showed two significant spectral breaks at \SI{304}{\MeV} and \SI{8.4}{\GeV} by combining the source spectrum from \textit{Fermi}-LAT, which was believed to be indicative of $\gamma$-ray emission produced through neutral-pion decay.

2. The expected significance of HESS J1912+101 observed by HADAR was \SI{20.6}{\sigma}. HESS J1912+101 (2HWC J1902+099) is associated with the high spin-down luminosity pulsar PSR J1913+1011 ($\dot{E} = \SI{2.9e36}{erg.s^{-1}}, d = \SI{4.6}{kpc}, \tau = \SI{169}{kyr}$)~\cite{J1913+1011 2008}. H.E.S.S. reported the integral flux above \SI{1}{TeV} was \SI{2.49 \pm 0.35 e-12}{cm^{-2}.s^{-1}} (11\% Crab flux) with spectral index \num{2.56\pm0.09}.  ARGO-YBJ also detected emission from this region~\cite{argo-ybj gamma survey 2013}, but the flux above \SI{1}{TeV} was much higher than the corresponding result of H.E.S.S.. This discrepancy was also present for other ARGO-YBJ sources and the reasons had been explained~\cite{argo J1841 2013}. HAWC also reported this source~\cite{HAWC 2017}, the photon index was \num{-2.64} and the differential energy spectrum was \SI{3.66e-14}{TeV^{-1}.cm^{-2}.s^{-1}} at \SI{7}{TeV}, which was used in this simulation.

3. The expected significance of W51C was \SI{8.2}{\sigma}. W51C (2HWC J1922+140, HESS J1923+141) is a radio-bright SNR located at a distance of $\sim$\SI{5.4}{kpc}~\cite{w51c distance 2018} and is a middle-aged remnant ($\sim$\SI{3e4}{yr})~\cite{w51c koo 1995}. W51C is known to interact with the molecular clouds in the star forming region W51B. The $\gamma$-ray emission from this region was discovered by \textit{Fermi}-LAT (\SI{200}{\MeV}--\SI{50}{\GeV})~\cite{w51c fermi 2009}, H.E.S.S. ($\textgreater\SI{1}{TeV}$)~\cite{w51c hess 2009}, MAGIC ($\sim$\SI{50}{\GeV}--several TeV)~\cite{w51c magic 2012}, and HAWC ($\sim$TeV)~\cite{HAWC 2017}. The $\gamma$-rays spectral energy distribution presented by \textit{Fermi} can be explained reasonably by the hadronic emission mechanism~\cite{w51c fermi 2009}. The MAGIC collaboration reported the integrated flux above \SI{1}{TeV} is equivalent to $\sim$3\% of the flux of the Crab Nebula~\cite{w51c magic 2012}, which agrees with the flux measurement by H.E.S.S., and the photon index is $\rm 2.58\,\pm\,0.07_{stat}\,\pm\,0.22_{sys}$. Above \SI{1}{TeV}, MAGIC shows the W51C region is composed of two components: one is coincident with the interaction region between W51C and W51B, while the other is coincident with the PWN CXO J192318.5+140305~\cite{w51c magic 2012}. HAWC observation result under the point source search agrees with the MAGIC and \textit{Fermi}-LAT results, whereas the extension is large~\cite{HAWC 2017}, and the extension with a radius of \ang{0.9} from HAWC is adopted in this simulation.  

4. The expected significance of ARGO J2031+4157 was \SI{23.6}{\sigma}. ARGO J2031+4157 is associated with TeV J2032+4130, a PWN first discovered as an unidentified source in TeV by HEGRA experiment~\cite{J2032 HEGRA 2005} in the Cygnus region, with no obvious counterpart at other wavelengths. In subsequent, Wipple~\cite{cygnus whipple 2007}, MAGIC~\cite{J2032 MAGIC 2008} and VERITAS~\cite{J2032 veritas} also reported this source, and the extended emission was about \ang{0.2} from the pulsar position, the integral flux above \SI{1}{TeV} was $\sim$3\%--6\% of the Crab Nebula flux. TeV J2032+4130 is associated with the PWN of PSR J2032+4127~\cite{PSR J2032+4127 binary 2018}. The Cygnus Cocoon is the GeV diffuse emission region, with a 2D Gaussian width of 2.0 $\pm$ \ang{0.2}, first reported by the \textit{Fermi} collaboration~\cite{cygnus fermi 2011}, and originates from a superbubble of freshly accelerated cosmic rays. The extended source ARGO J2031+4157~\cite{argo-ybj cygnus 2012}, MGRO J2031+41~\cite{galactic milagro 2007}, VER J2031+415~\cite{J2032 veritas}, 2HWC J2031+415~\cite{HAWC 2017}, and LHAASO J2032+4102~\cite{1.4PeV 2021} were reported compatible with the Cygnus Cocoon, and they were consistent in flux relatively. A source extension of \ang{2} was used in our simulation concerning the results of previous measurements. In HADAR FOV, the effective observation time for Cygnus Cocoon is 277.7 hrs in 1 yr operation time with zenith angle between \ang{12.4} and \ang{30}.  

5. The expected significance of Cassiopeia A was \SI{5.7}{\sigma}. Cassiopeia A is the youngest ($\sim$\SI{350}{yrs}~\cite{Cas 350yr 1980}) SNR of the historical galactic SNRs, located at a distance of \SI{3.4}{kpc}~\cite{Cas 3.4kpc 1995}. It is bright in radio and X-rays and has been detected as a bright point source in MeV--GeV $\gamma$-rays by \textit{Fermi}-LAT and in VHE $\gamma$-rays by HEGRA, MAGIC and VERITAS. The observations of \textit{Fermi}-LAT~\cite{fermi cassiopeia 2010} and VERITAS on Cassiopeia A in MeV to GeV energy range have shown that the hadronic acceleration model is preferred to fit the energy spectrum~\cite{Fermi cassiopeia break yuan 2013}. HEGRA first detected TeV $\gamma$-ray emission associated with Cassiopeia A above \SI{1}{TeV}~\cite{cassiopeia evidence HEGRA 2001}, with later confirmations supplied by MAGIC~\cite{cassiopeia MAGIC 2007} and VERITAS~\cite{cassiopeia veritas 2010}. The spectral index (\numrange[range-phrase = --]{2.4}{2.6}) and the fluxes (3\% of the Crab Nebula flux) reported by these groups are consistent with each other within errors. The results also indicated that a strong magnetic field was needed to produce the TeV $\gamma$-ray emission. The zenith angle for Cassiopeia A in HADAR FOV was \SIrange[range-phrase = --]{28.7}{30}{\degree}, thus the effective observation time was relatively short. The reference differential energy spectrum used in the simulation was the result of VERITAS, and the spectral index was 2.75.    

\subsection{Expected Observations of PWNs}\label{4.2}
1. The expected significance of Crab Nebula was the largest significance of the TeV sources, with \SI{346.0}{\sigma}. The Crab nebula is a leftover of the supernova explosion and is powered by the pulsar PSR B0531+21 at its center~\cite{crab PSR B0531 2008}. It is the brightest steady source in the northern sky, and has been measured precisely with IACTs and ground-based experiments over the past three decades~\cite{crab 1989,crab hess 2006,crab HEGRA 2004,crab VERITAS icrc 2015,crab MAGIC 30 2015,crab HAWC 2017}. In the GeV regime, the synchrotron radiation is observed from radio to soft $\gamma$-rays. At higher energies (\textgreater \SI{1}{TeV}), the Inverse Compton up-scattering by the relativistic electrons accelerated in shocks in the nebula is dominant~\cite{crab gamma radia 1996}. Due to the steady brightness, Crab Nebula has been adopted as the reference source in TeV astronmy and used for calibrating new TeV $\gamma$-ray instruments. A joint fit of the Crab Nebula spectrum was produced by combining data from \textit{Fermi}-LAT and the four current operating IACTs, and the normalization flux at \SI{1}{TeV} was \SI{3.85\pm0.11e-11}{TeV^{-1}.cm^{-2}.s^{-1}} and the spectral index was \num{2.51 \pm 0.03}~\cite{crab joint fit 2019}.    

2. The expected significance of Geminga was \SI{13.7}{\sigma}. Geminga is the first known radio-quiet pulsar and the second brightest source in the GeV sky. It is a promising candidate for the emission of VHE pulsed $\gamma$-rays. Geminga (PSR J0633+1746) is relatively old (\SI{342}{yrs}), nearby ($250_{-62}^{+120}$ pc) and has a low spin-down power (\SI{3.26e34}{erg.s^{-1}})~\cite{geminga 1992,geminga neutron star 2007}. Detection of extended $\gamma$-ray emission around Geminga was discovered by Milagro~\cite{geminga milagro 2009} and HAWC. The angular size reported for the source was $\sim$\ang{2}, challenging for IACTs to detect due to the large scale emission~\cite{geminga magic 2016}. HAWC confirmed the Milagro detection of VHE $\gamma$-ray emission around Geminga and another nearby pulsar PSR B0656+14~\cite{HAWC 2017}. In 2017, HAWC detected a TeV $\gamma$-ray halo around Geminga. Using the observed surface brightness distribution, they determined that the diffusion coefficient of electrons and positrons is two orders of magnitude lower than that derived from the typical interstellar medium, thus constraining the origin of the positron flux at Earth~\cite{geminga hawc 2017,geminga liury 2019,geminga constrain hawc 2018,fangkun 2018}.  

3. The expected significance of TeV J1930+188 was \SI{10.5}{\sigma}. The VHE $\gamma$-ray source TeV J1930+188 (VER J1930+188, 2HWC J1930+188, HESS J1930+188) is associated with the composite SNR G54.1+0.3 and the pulsar PSR J1930+1852 ($\dot{E} = \SI{1.2e37}{erg.s^{-1}}, d = \SI{6.5}{kpc}, \tau = \SI{2.9}{kyr}$)~\cite{G54.1+0.3 rado and x-ray 2002}. VERITAS first discovered the TeV source, and the observation is consistent with a point-like source~\cite{J1930+188 VERITAS 2010}. The spectral index at \SI{7}{TeV} of VERITAS, HAWC~\cite{HAWC 2017} and H.E.S.S.~\cite{Hess survery 2018} are consistent with statistical and systematic uncertainties. Whereas, the integral flux above \SI{1}{TeV} reported by H.E.S.S. is lower than that of VERITAS. The detailed observation comparison between VERITAS and HAWC for the SNR G54.1+0.3 was given in~\cite{VERITAS Fermi HAWC 2018}, and the VERITAS flux for G54.1+0.3 is lower than that measured by HAWC.  

4. The expected significance of 2HWC J1953+294 was \SI{31.4}{\sigma}. 2HWC J1953+294 was first discovered by HAWC in 2017~\cite{HAWC 2017}, after that, VERITAS reported a confirmation of $\gamma$-ray emission coincident with 2HWC J1953+294 with an extended-source analysis (\ang{0.3}) with 37 hr and 72 hr respectively (named VER J1952+293)~\cite{J1953+294 VERITAS 2017}. The counterpart of 2HWC J1953+294 (VER J1952+293) is the PWN DA 495 (age $\sim$\SI[group-separator = {,}]{20000}{yr}, distance \SI{1.0}{kpc})~\cite{DA 495 2008}, which is associated with the SNR G65.7+1.2. The spectral index measured by VERITAS is $\rm 2.65\,\pm\,0.49_{stat}$, which is consistent with the HAWC result $\rm 2.78\,\pm\,0.15_{stat}$ within errors. However, the VERITAS flux for DA 495 at \SI{1}{TeV} is lower than that measured by HAWC, it is mainly due to the different search modes for this source (\ang{0.3} extension search or point-like source search)~\cite{VERITAS Fermi HAWC 2018,HAWC 2017}. The flux and spectral index adopted in our simulation is from HAWC.     

\subsection{Expected Observations of Gamma-Ray Binary}\label{4.3}
The known $\gamma$-ray binaries are usually identified as high-mass X-ray binaries system, consisting of a compact object orbiting around a massive stellar of O or Be type. The class of TeV binaries is quite sparse, only consisting of PSR B1259-63~\cite{B1259-63 2005}, LS 5039~\cite{LS 5039 2005}, LS I +\ang{61}303~\cite{ls i+ 61 303 2006,ls i+ 61 303 2016}, HESS J0632+057~\cite{HESS J0632+057 2007}, 1FGL J1018.6-5856~\cite{1fgl j1018.6-5856 2015}, and PSR J2032+4127~\cite{PSR J2032+4127 lyne 2015,PSR J2032+4127 Ho 2017,PSR J2032+4127 binary 2018}, and every member shows different characteristics. The IACTs are qualified to study these objects which exhibit the TeV emission. Of these binaries systems, most of the nature of the compact object for systems are unknown, except for the compact objects of PSR B1259-63 and PSR J2032+4127, which have been firmly identified as pulsars~\cite{Psr 1259-63-a binary radio pulsar 1992,gamma pulsars fermi 2009}. Consequently, the fundamental mechanism for the TeV emission remains uncertain. The orbital periods of these systems vary from several days (LS 5039) to many years (PSR J2032+4127, approximately 50 years~\cite{PSR J2032+4127 binary 2018}). The TeV emission varies obviously as a function of the orbital phase, thus the flux variations may provide short windows for the instruments to detect in the TeV regime. 

In the northern hemisphere, only the positions of HESS J0632+057, PSR J2032+4127, and LS I +\ang{61}303 in the sky are allowed to be observed, furthermore, only HESS J0632+057 and PSR J2032+4127 locate in the HADAR FOV. PSR J2032+4127 is spatially coincident with source TeV J2032+4130 which has been described as an extended VHE source in section~\ref{4.1}.   

Gamma-ray emission at VHE from HESS J0632+057 was first discovered serendipitously at energies above \SI{400}{\GeV} with H.E.S.S. experiment~\cite{HESS J0632+057 2007}. This $\gamma$-ray source is coincident with the massive B-star MWC 148 at a distance of \numrange[range-phrase = --]{1.1}{1.7} kpc~\cite{blue giant star 1955,J0632+057 a new gamma binary?,J0632+057 optical 2010}. It was observed in the following years with VERITAS, H.E.S.S., and MAGIC telescopes~\cite{hess j0632+ 057 2009,J0632+057,J0632+057 MAGIC 2012}. In 2010 and 2011, significant $\gamma$-ray signals were detected at energies above \SI{1}{\TeV}, confirming the TeV variability. With the measurement of flux modulations, the period of 317.3 $\pm$ 0.7 days was determined with the X-ray observations~\cite{hess j0632+ 057 period 2011,hess j0632+ 057 period 2009}, and the binary nature of HESS J0632+057 was firmly established~\cite{j0632+057 binary nature}. According to the $\gamma$-ray light curve analysis from the latest data, the two higher flux states occurred in the orbital phase range 0.2--0.4 and 0.6--0.8, the low state in the phase range 0.4--0.6, and the medium (plateau) state in the remaining ranges 0.8--0.2. The detailed spectral analyses at $\gamma$-ray energies averaged over the four different ranges in orbital phase were shown in~\cite{hess j0632+057 2021}. On the basis of the medium state data observed by VERITAS, the flux normalization at \SI{0.5}{\TeV} is \SI{20.8 \pm 3.1 e-13}{TeV^{-1}.cm^{-2}.s^{-1}}, and the photon index is 2.67 $\pm$ 0.17~\cite{hess j0632+057 2021}, which are used in the calculation for HADAR. The expected significance of HESS J0632+057 binary was \SI{6.0}{\sigma} in 5 yr operation time, the weak flux in medium state contributes to the low detection significance, thus the significance is not outstanding. With the continual observation of the binaries for a long time, HADAR may provide more clues to investigate the $\gamma$-ray binaries.

\subsection{Expected Observations of Unidentified Sources}\label{4.4}
For the 14 unidentified sources that we expected to be observed by HADAR, none of them had associations with known physical objects. Most of these sources were discovered by HAWC, which benefited from its sensitivity above TeV. HADAR is sensitive to $\gamma$-rays in the energy range from \SI{10}{\GeV} to tens of TeV, and its sensitivity is similar to the H.E.S.S. and MAGIC experiments in the TeV band. Moreover, HADAR also compensates for the weakness of HAWC and LHAASO observations below the TeV level. Therefore, when the HADAR experiment becomes operational, more identifications or associations of these unidentified sources will be provided. Also, we expect to discover more new sources emitting GeV--TeV $\gamma$-ray emissions.

\subsection{Expected Observations of \textit{Fermi}-LAT Sources}\label{4.5}
To further evaluate the ability of observations in low energy bands, we also analyzed the potential of HADAR for sky surveys in the GeV energy band based on the \textit{Fermi}-LAT 4FGL sources~\cite{Fermi 4fgl 2020}. The 4FGL catalog covers the \SI{0.05}{\GeV} to \SI{1}{\TeV} energy range, which partially overlaps with the \SI{30}{\GeV} to \SI{1}{\TeV} range of HADAR. Here, we pay particular attention to the energy range below \SI{100}{\GeV}. The one-year integral sensitivity in this energy range of HADAR is $\sim$4.8\% of the flux from the Crab Nebula, which is better than the \textit{Fermi}-LAT. In terms of the fitting parameters of the spectrum provided by 4FGL, we studied the observed statistical significance of different types of sources in the HADAR FOV. The results, listed in Table~\ref{tab5}, show a total of 506 sources from 4FGL located within the HADAR region, excluding extragalactic sources. We found that 62 of the 506 sources were observed by HADAR in \SI{1}{yr} of operation. Most of the sources in 4FGL could not be observed, mainly due to the low flux or the broken power-law at low energies. Of the observed types, it should be noted that pulsars only represent indirect sources. Pulsars do not directly emit unpulsed $\gamma$-ray emissions, but rather indicate that they could be powering a PWN that directly discharges such emissions.

It is also remarkable that the limited size of the \textit{Fermi}-LAT detector yielded a reduced efficiency in the VHE domain (TeVCat) survey. The number of 4FGL source associations listed in TeVCat is only 57~\cite{Fermi 4fgl 2020} and these sources often have harder spectra or higher break energies. Due to its higher sensitivity below \SI{100}{\GeV}, HADAR is expected to detect more $\gamma$-ray sources, including sources that emit $\gamma$-rays following a spectrum that extends from the GeV to the TeV range.  

\begin{table*}[!htb]
	\caption{Expected observation numbers of galactic $\gamma$-ray sources as well as unknown and unassociated sources in the \textit{Fermi}-LAT 4FGL catalog with HADAR below \SI{100}{\GeV}. The third column lists the number of objects in the HADAR survey regions, while the fourth column provides the number of sources expected to be observed by HADAR.}
	\label{tab5}
	\tabcolsep 6.0pt
	\renewcommand{\arraystretch}{1.2} 
	\begin{threeparttable}
		\begin{tabular*}{\textwidth}{c c c c} 
			\toprule
			\hline
			Type & \tabincell{c}{Number of sources in \\4FGL} & \tabincell{c}{Number of sources in HADAR \\region} & \tabincell{c}{Expected to be observed by \\HADAR}\\ \hline
			Pulsars & 239 & 86 & 36 \\
			PWNe, SNR, Star-forming region & 61 & 21 & 12 \\ 
			SNR/PWNe (SPPs) & 78 & 18 & 0 \\
			Globular cluster & 30 & 4 & 0 \\
			\tabincell{c}{High-mass Binary, Low-mass Binary, \\Binary, Nova} & 12 & 4 & 0\\ \hline
			Unknown & 92 & 27 & 3 \\
			Unassociated & 1336 & 346 & 11 \\ \hline
			Total & 1848 & 506 & 62\\ 
			\bottomrule
			\hline
		\end{tabular*}
		\begin{tablenotes}
			\footnotesize
			\item\textbf{Notes.}
			\item[1] Extragalactic source types are not included in this table and the unassociated sources also statistically consist of some types of extragalactic sources. 
			\item[2] "SPPs" indicates potential association with SNRs or PWNe.
		\end{tablenotes}
	\end{threeparttable}
\end{table*}

\subsection{Predicted Observations of the Extended Sources}\label{4.6}
Concerning realistic observations, hundreds of extended sources have been revealed in the GeV to multi-TeV energy range~\cite{Hess survery 2018,Fermi 4fgl 2020,3HWC 2020}. Particle detectors such as HAWC and LHAASO have discovered increasingly extended TeV sources around middle-aged pulsars. Many of them possess halo structures and are likely to be in the halo evolutionary phase~\cite{geminga hawc 2017}. Therefore, it is vital to investigate the radial distribution and morphology of extended $\gamma$-ray sources. For instance, the surface brightness profile provides important information regarding cosmic ray injection and is a useful tool for determining the radiation mechanism and propagation process of parent relativistic particles~\cite{yangrz radial profile 2022}. However, there are technical challenges involved in the observation of highly extended $\gamma$-ray emissions using IACTs. Taking the Geminga pulsar as an example, the emissions around it are considerably extended, with a $\sim$\SI{5.5}{\degree} radius in the 8--\SI{40}{\TeV} energy range, as measured by HAWC~\cite{geminga hawc 2017}. This full scale of emissions is beyond the range of the typical IACT FOV ($\sim$3--\SI{5}{\degree}), making detection particularly challenging. Besides, background estimations remain difficult because in the standard background approach the background is estimated from regions that are free of source emissions. However, sources often fill the FOV or there are very few remaining source-free regions, such that the source-free regions are insufficient for background estimation. Although some new background methods have been proposed~\cite{geminga hess 2021}, the precision of the background estimation is also essential for correctly investigating the emission morphology of the extended sources.  

HADAR owns the ability to continuously survey the sky with a large FOV, thus it can effectively study extremely extended sources such as Geminga and ARGO J2031+4157. Meanwhile, HADAR could observes abundant source-free regions for background estimation. It should be noted that to investigate the profile of extended $\gamma$-ray sources, the PSF of instruments plays a significant role. For the VHE domain (around \SI{10}{TeV}), the angular resolution of HADAR is approximately \SI{0.25}{\degree}, which is smaller than with IACTs. Thus, there may be some limitations regarding the spatial distribution analysis of extended sources.  

\section{Conclusions}\label{section5}
For two decades, the ability to observe VHE gamma-ray sources has greatly improved. We have benefitted from the construction of large ground-based observatories using either the IACT approach (such as H.E.S.S., MAGIC, VERITAS, and the upcoming CTA) or an EAS detector array (such as HAWC and LHAASO). IACT observes the VHE sky with excellent sensitivity and angular resolution ($\sim$\SI{0.1}{\degree}), but it is strictly limited to performing regular monitoring of sources, mapping vast regions of the sky, and promptly responding to fast transients due to its narrow FOV and its duty cycle.  Observational coverage and continuity problems could be overcome using the wide FOV of EAS facilities. However, the energy threshold is relatively high (hundreds of GeV) and the angular resolution is relatively poor ($\sim$\SI{0.3}{\degree}). The proposed HADAR experiment possesses excellent sensitivity, low-energy threshold, and wide FOV.

HADAR is a new atmospheric Cherenkov telescope experimental array that employs a novel observational technique using large-dimensional refracting water-lenses to focus the atmospheric Cherenkov light generated in air showers. With a wide FOV several times that of the reflective atmospheric Cherenkov telescope, HADAR could continuously surveys and monitors the sky of VHE $\gamma$-ray sources. In this paper, we estimated the significance of TeV $\gamma$-ray sources in the galaxy from \SI{30}{\GeV} to \SI{10}{\TeV} over a one year operation time.

According to the expected results, a total of 23 point sources in TeVCat were detected with more than \SI{5}{\sigma} significance, including five SNRs and superbubbles, four PWNs, and 14 unidentified sources. Furthermore, we briefly discussed the measurement results of some previous experiments. In terms of the significance results for the Crab Nebula, the statistical significance is expected to be \SI{346.0}{\sigma}, while the one-year integral sensitivity of HADAR above \SI{1}{TeV} is $\sim$1.3\%--2.4\% of the flux from the Crab Nebula. With such high sensitivity, especially in the energy range of \SIrange[range-phrase = --,range-units = single]{30}{100}{\GeV}, HADAR will link the energy ranges of the \textit{Fermi}-LAT and H.E.S.S. experiments. This will offer great improvements in energy spectrum measurement for sources in the GeV band.

Regarding the HADAR experiment, a water-lens telescope array of \SI{1}{\m} diameter will be installed and operational in late 2022. At that time, the HADAR telescope will provide more experience in the exploration of imaging, triggering, event reconstruction, and $\rm \gamma/Proton$ separation. Moreover, the manufacturing process of the \SI{5}{\m} diameter water-lens telescope is currently in testing. As the HADAR experiment continues, more $\gamma$-ray sources with GeV--TeV emissions will be discovered due to its high sensitivity, thereby allowing us to investigate many more $\gamma$-ray sources in the galaxy.

\acknowledgements{%
	We acknowledge the support from the National Natural Science Foundation of China (Nos. 11873005, 11705103, 12005120, 12147218, U1831208, U1632104, 11875264, and
	U2031110).
}

\end{document}